\documentclass[%
 reprint,
superscriptaddress,
amsmath,amssymb,
aps,
pre,
showpacs,
longbibliography
]{revtex4-1}

\usepackage[utf8]{inputenc}
\usepackage{graphicx}
\usepackage{dcolumn}
\usepackage{bm}
\usepackage{tikz}

\usepackage{pgfplots}
\pgfplotsset{compat=1.3}
\usetikzlibrary{plotmarks}

\newcommand{\usv}{^{\rm loc}}
\newcommand{\rhe}{^{\rm rheo}}

\begin{document}
\title{Local Oscillatory Rheology from Echography}

\author{Brice Saint-Michel}
 \email{brice.saint-michel@ens-lyon.fr}
\author{Thomas Gibaud}
 \email{Corresponding author, thomas.gibaud@ens-lyon.fr}
 \affiliation{Laboratoire de Physique, Universit\'e de Lyon, \'Ecole Normale Sup\'erieure de Lyon, CNRS UMR 5672,  \\ 
	46 All\'ee d’Italie, 69364 Lyon cedex 07, France }%
\author{Mathieu Leocmach}
 \email{mathieu.leocmach@univ-lyon1.fr}
 \affiliation{Institut Lumi\`ere Mati\`ere, CNRS UMR 5306, Universit\'e Claude Bernard Lyon 1, Universit\'e de Lyon, Lyon, 69622 Villeurbanne Cedex, France}%
\author{S\'ebastien Manneville}
	\email{Corresponding author, sebastien.manneville@ens-lyon.fr}
 \affiliation{Laboratoire de Physique, Universit\'e de Lyon, \'Ecole Normale Sup\'erieure de Lyon, CNRS UMR 5672,  \\ 
	46 All\'ee d’Italie, 69364 Lyon cedex 07, France }%

\date{\today}

\begin{abstract}
Local Oscillatory Rheology from Echography (LORE) consists in a traditional rheology experiment synchronized with high-frequency ultrasonic imaging which gives access to the local material response to oscillatory shear. Besides classical global rheological quantities, this method provides quantitative time-resolved information on the local displacement across the entire gap of the rheometer. From the local displacement response, we compute and decompose the local strain in its Fourier components and measure the spatially-resolved viscoelastic moduli. After benchmarking our method on homogeneous Newtonian fluids and soft solids, we demonstrate that this technique is well suited to characterize spatially heterogeneous samples, wall slip, and the emergence of nonlinearity under large amplitude oscillatory stress in soft materials.
\end{abstract}


\maketitle


\section{Introduction}
A classical oscillatory stress experiment performed with a rheometer consists in applying a sinusoidal oscillatory stress $\sigma\rhe(t)$ at frequency $f$ and amplitude $\sigma_1$ on a soft material to test its strain response $\gamma\rhe(t)$. If the sample is a purely homogeneous viscous fluid, the stress is proportional to the shear rate, i.e. the time-derivative of the strain, $\sigma\rhe(t)=\eta\dot\gamma\rhe(t)$, with $\eta$ the viscosity of the fluid. In contrast, if the sample is a homogeneous elastic solid, then the stress amplitude is proportional and in phase with the strain, $\sigma\rhe(t)=G_0\gamma\rhe(t)$, with $G_0$ the elastic modulus of the material~\cite{Macosko1994,Larson1998,Russell1989,Chen2010}. Most soft materials, e.g. gels, viscoelastic fluids, suspensions and pastes, fall between these two ideal cases, exhibiting a phase lag between $\sigma\rhe(t)$ and $\gamma\rhe(t)$ that ranges from $0$ to $\pi/2$ and depends on the oscillation frequency $f$.

Standard rheological analyses rely upon the basic assumption that the sample behaves homogeneously across the gap of the shear geometry. Yet many materials become spatially heterogeneous under shear and classical rheology, providing only averaged information over the whole sample, fails in capturing such spatial heterogeneities. Examples include shear-banding instabilities, shear localization, fractures within the material and apparent wall slip. Shear banding is observed in viscoelastic fluids such as wormlike micellar solutions~\cite{Berret2006, Lerouge2010} and telechelic polymers~\cite{Manneville2007,Sprakel2008}. Within a given range of shear rates, the flow segregates into macroscopic bands with different local viscosities and stacked along the velocity gradient direction. A particular case of shear banding, sometimes referred to as shear localization, is observed close to the yielding transition of viscoelastic solids, such as colloidal gels~\cite{Coussot2002, Divoux2012, Moller2006}, star polymers~\cite{Rogers2008}, emulsions~\cite{Becu2006} and foams~\cite{Gilbreth2006}. Here only some part of the sample flows while the rest remains solid. Soft solids~\cite{Baumberger2006,Leocmach2014} and viscoelastic fluids~\cite{Pignon1996,Skrzeszewska2010,Ligoure2013} may also display bulk fracture when stressed deep into the nonlinear regime, associated either to irreversible failure in the former case or to self-healing  mechanisms in the latter case. Finally, apparent wall slip is observed ubiquitously in complex fluids, especially under smooth boundary conditions \cite{Yoshimura1988,Barnes1995}. Characterizing and understanding wall slip at the microscale stand out as experimental and theoretical challenges both in the case of partial wall slip (where the sample is sheared in the bulk with a local shear rate smaller than the global shear rate)~\cite{Lettinga2009,Divoux2015arXiv} and in the case of total wall slip (where the sample displays a pluglike flow, i.e. solid-body rotation, in the entire gap of the rheometer)~\cite{Meeker2004, Gibaud2009, Ballesta2012, Seth2012}. In all cases above, the global rheological response does not correctly reflect the material properties and a local investigation is mandatory to assess the actual sample properties.

Several local techniques have been developed to examine the local rheology of soft materials. These techniques are based on birefringence~\cite{Lerouge1998}, light scattering \cite{Boltenhagen1997,Salmon2003}, optical microscopy~\cite{Cohen2006,Besseling2007,Nordstrom2010,Jop2012}, magnetic resonance imaging~\cite{Callaghan1999, Coussot2002} or ultrasonic velocimetry~\cite{Manneville2008, Gallot2013}. So far these methods have been mainly used to probe creep or flow experiments and only a few studies, all based on optical particle tracking, have focused on the case of time-resolved local measurements under oscillatory shear~\cite{Tapadia2006,Boukany2008,Li2009,Guo2011,Dimitriou2012,Li2013}. Among them, only Guo {\it et al.} \cite{Guo2011} have reported on local viscoelastic measurements under large-amplitude oscillatory shear (LAOS). However these measurements only focused on estimates of the local strain amplitude in granular suspensions, without any analysis of the phase information or of the harmonic content of the local data. 

Here we introduce a technique referred to as Local Oscillatory Rheology from Echography (LORE) which consists in synchronizing high-frequency ultrasonic imaging with standard oscillatory stress rheology. Building upon our previous work, which was restricted either to steady flows~\cite{Gallot2013} or to stroboscopic measurements under LAOS~\cite{Perge2014}, we show that ultrasonic imaging is particularly well-suited for time-resolved local measurements during oscillatory experiments. Indeed the acquisition frequency of up to 20,000~fps is fast enough to capture local displacements $\Delta\usv$ under oscillations with frequencies up to 1~kHz and strain amplitudes down to a few percents. It also offers the possibility to probe optically opaque materials. We further demonstrate that LORE provides full access to the local strain $\gamma\usv$ and to the local elastic and viscous moduli, $G'_{\rm loc}$ and $G''_{\rm loc}$, as well as to higher frequency components in the case of nonlinear material response.

The paper is organized as follows. In Sect.~\ref{matmeth} we present the specifications of the apparatus combining standard rheology and ultrasonic imaging. Then we introduce the method used to map the local viscoelastic moduli. In Sect.~\ref{newtonian} we benchmark the LORE technique using a homogeneous Newtonian fluid and a soft elastic solid. We demonstrate that our method successfully resolves the local viscoelastic moduli of the material across the entire 2-mm gap of a Taylor-Couette cell and matches the rheometer average measurements. We then verify in Sect.~\ref{hetero} that LORE provides access to the local values of the elastic modulus in a spatially heterogeneous soft solid. In particular we show that the displacements are confined to the softest region of the material. In Sect.~\ref{slip} we finally examine the influence of slippery boundary conditions on the harmonic content of the local strain response of a soft solid. 


\section{Materials and methods}
\label{matmeth}

\subsection{Rheological measurements}

We apply an oscillatory shear stress to fluids or soft solids using a commercial stress-imposed rheometer (TA Instruments AR G2). Our rheological experiments are performed in a homemade Taylor-Couette cell with smooth, polymethylmethacrylate (PMMA) walls. The inner rotating cylinder (rotor) has a radius $R_{\rm i} = 23$~mm, a height $H = 6$~cm. Its upper part is attached to the rheometer and its bottom part is terminated by a cone with an angle of $2^\circ$ with a truncation of 50~$\mu$m. The fixed outer cylinder (stator) has an inner radius $R_{\rm o}= 25$~mm. The temperature is controlled by a water circulation around the Taylor-Couette cell and fixed to $25\pm0.1^\circ$C for all experiments. The sample is introduced in the radial gap $e = 2$~mm between the rotor and the stator and submitted to an oscillatory shear stress $\sigma\rhe(t) = \sigma_1 \cos(2 \pi f t)$ with frequency $f = 0.1$~Hz and amplitude $\sigma_1$ [see Fig.~\ref{fig:setup}(a)].

\subsection{Ultrasonic imaging under oscillatory stress}

We synchronize our rheological measurements with high-frequency ultrasonic imaging~\cite{Gallot2013}, a technique that is fast enough to reconstruct time-resolved local displacements within the sheared sample during oscillatory stress experiments. To provide ultrasonic contrast, the samples are seeded with density-matched passive tracers (Potters Sphericel\textregistered{} 110P8 hollow glass microspheres of median diameter $D_{50} \simeq 10~\mu$m and density $d = 1.10$, or Arkema Orgasol polyamide particles, grade 2002 ES3 NAT3, with $D_{50} \simeq 30~\mu$m and density $d = 1.03$). These tracers are almost density-matched with the suspending medium and their concentration of 1 to 3 wt.~\% is high enough to obtain sufficient ultrasonic scattering yet low enough to prevent multiple scattering and to ensure that they do not affect the sample mechanical properties.

\begin{figure}
	\centering
	\includegraphics{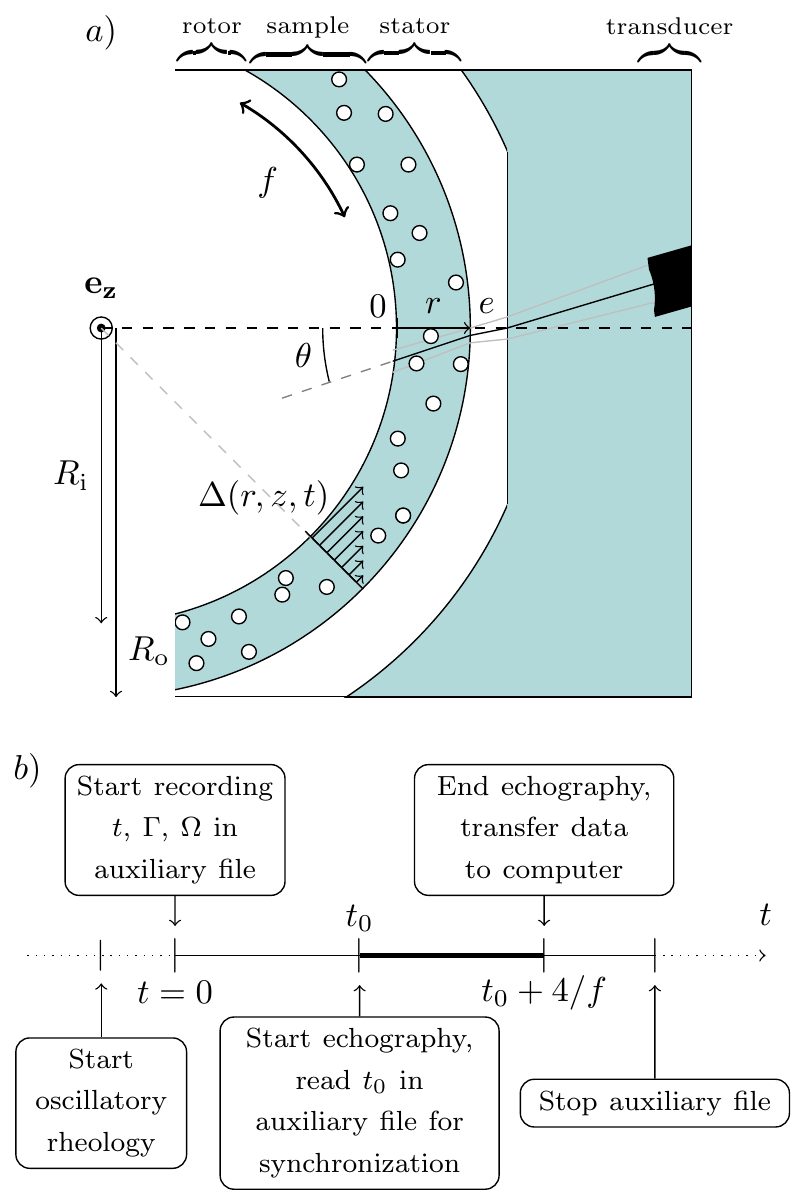} 
    \caption{(a) Experimental setup for LORE measurements. (b)~Block diagram describing the typical experimental work flow.}
   \label{fig:setup}
\end{figure}

Our high-frequency ultrasonic imaging device relies on a linear array of 128 piezoelectric transducers with a total active length of 32~mm. As already described at length by Gallot {\it et al.}~\cite{Gallot2013}, the transducer array is immersed in the water tank surrounding the Taylor-Couette device and is set at about 25~mm from the stator. As sketched in Fig.~\ref{fig:setup}(a), short ultrasonic plane pulses with a central frequency of 15~MHz impinge on the stator and propagate across the gap at an angle $\theta\simeq 5^\circ$ relative to the radial direction ${\bf e_r}$. These pulses get scattered by the seeding particles within the sample and the backscattered signal is recorded by the transducer array, leading to an ``ultrasonic speckle'' signal with 128 measurement lines and typically 640 points sampled at 160~MHz. The analysis of ultrasonic data consists in first processing the speckle signal into a beam-formed speckle image $S(r, z, t)$, where $r$ is the radial distance across the gap (${\bf e_r}$ points outward with $r=0$ taken at the rotor and $r=2$~mm at the stator) and $z$ is the vertical direction (${\bf e_r}$ points downward with $z=0$ taken at about 10~mm from the top of the Taylor-Couette device). Then two successive speckle images are cross-correlated in order to get the tangential displacement $\Delta\usv$ as a function of $r$ and $z$. Here we set the time interval between two speckle images to $1/(600 f)$ and the total acquisition time to $4/f$. This allows us to resolve the displacement $\Delta\usv(r, z, t)$ during four oscillation periods with a spatial resolution along the $z$-direction of 250~$\mu$m and 75~$\mu$m in the radial direction $r$ and with a sampling of 600 images per oscillation period. To increase the signal-to-noise ratio, displacement maps are further averaged over 10 successive cross-correlations, which provides a time resolution of $1/(60 f)$ (see Supplemental Movie~1).

\subsection{LORE data analysis}
\label{corrections}

By combining ultrasonic imaging and rheology, we reconstruct local rheological quantities such as the strain and the viscoelastic moduli. A first step is to compute the local shear stress within the gap of our Taylor-Couette device. For oscillatory stress experiments, the rheometer applies a raw torque $\Gamma(t)$ and monitors the raw rotor angular velocity $\Omega(t)$. From the momentum conservation equation in cylindrical coordinates we obtain the local stress across the gap~\cite{Macosko1994,Perge2014,Baravian2007}:
\begin{equation}
	\sigma\rhe(r,t) = \frac{\Gamma(t) - J \dot{\Omega}(t)}{2 \pi H (R_{\rm i} + r)^2} \,.
    \label{eq:sigma_r_bis}
\end{equation}
Equation~\eqref{eq:sigma_r_bis} indicates that the stress varies across the gap due to the curvature of the cylindrical geometry. In our particular conditions, $e/R_{\rm i}=0.087$ and the stress decreases by 18~\% from the rotor to the stator. The term $J \dot{\Omega}(t)$ in Eq.~\eqref{eq:sigma_r_bis} is a correction due to the inertia of the rotor. It depends on the momentum of inertia $J =50~\mu$N$\cdot$m$\cdot$s$^2$ of the rotor and on its acceleration. In the present experiments performed at $f=0.1$~Hz, this correction corresponds to at most 1~\% of the stress, which allows us to neglect inertia in Eq.~\eqref{eq:sigma_r_bis}. We have also checked that the harmonic modes of $\Gamma(t)$ are always negligible compared to the fundamental mode so that we can identify  $\sigma\rhe(r,t)$ with a pure cosine wave:
\begin{equation}
\sigma\rhe(r,t) = \sigma\rhe_1(r) \cos(2 \pi f t) \,,
\label{eq:strain_rheo_def2}
\end{equation}
where the local amplitude $\sigma\rhe_1(r)=\Gamma_1/[2 \pi H (R_{\rm i} + r)^2]$ is deduced from the amplitude $\Gamma_1$ of  the torque $\Gamma(t)$.

In a second step, we use the local tangential displacement $\Delta\usv(r, z, t)$ inferred from ultrasonic imaging to compute the local strain $\gamma\usv(r,t)$. In the present work, since the samples under scrutiny always remain homogeneous along the vertical direction, we use an average over the $z$-direction to improve the statistics:
\begin{equation}
	\gamma\usv(r,t) \equiv \Bigg \langle (R_{\rm i}+r)\, \partial_r \left(\frac{\Delta\usv(r,z,t)}{R_{\rm i}+r}\right) \Bigg \rangle_z
\label{eq:gamma_usv_def1}\\
\end{equation}
Homogeneity along the $z$-direction can be directly checked in Fig.~\ref{fig:z_homog} as well as in Supplementary Movie~1. We note however that such a $z$-average is not mandatory and that the LORE technique may also provide information resolved along the vertical direction if necessary. 

Moreover the strain response of the sample may be nonlinear in contrast to the stress input given by Eq.~\eqref{eq:strain_rheo_def2} \cite{Hyun2011}. This results in the presence of harmonics in the Fourier series decomposition of the local strain $\gamma\usv(r,t)$ which Fourier coefficients $\gamma_k\usv$ and phase lag $\phi_k\usv$ with respect to $\sigma\rhe(r,t)$ depend on $r$:
\begin{equation}
	\gamma\usv(r,t) = \sum_k \gamma_k\usv(r) \cos \left ( 2 k \pi f t + \phi_k\usv (r) \right ) \,.
\label{eq:gamma_usv_def2}
\end{equation}

Finally, based on the local shear stress $\sigma\rhe(r,t)$ given by Eq.~\eqref{eq:sigma_r_bis} and on the {\it fundamental} component of the local strain $\gamma\usv(r,t)$ of amplitude $\gamma_1\usv(r)$ and phase $\phi_1\usv(r)$, we define the {\it local} elastic and viscous moduli $G'_{\rm loc}$ and $G''_{\rm loc}$ respectively as:
\begin{align}
G'_{\rm loc}(r)  &= \frac{\sigma\rhe_1(r)}{\gamma_1\usv(r)} \cos(\phi_1\usv(r))\,, \\
G''_{\rm loc}(r) &= \frac{\sigma\rhe_1(r)}{\gamma_1\usv(r)} \sin(\phi_1\usv(r))\,.
\label{eq:G_usv}
\end{align}

These measurements of the local viscoelastic properties of the sample can then be easily compared to their global counterparts, namely the classical elastic and viscous moduli, $G'_{\rm rheo}$ and $G''_{\rm rheo}$, provided by the rheometer and defined by: 
 \begin{align}
G'_{\rm rheo} &= \frac{\sigma\rhe_1}{\gamma\rhe_1} \cos(\phi\rhe_1)\,, \\
G''_{\rm rheo} &= \frac{\sigma\rhe_1}{\gamma\rhe_1} \sin(\phi\rhe_1)\,,
\label{eq:G_rheo}
\end{align}
where $\sigma\rhe_1$, $\gamma\rhe_1$ and $\phi\rhe_1$ correspond to the fundamental Fourier component of the global rheological stress $\sigma\rhe(t)$ and strain $\gamma\rhe(t)$ measured by the rheometer:
\begin{align}
\sigma\rhe(t) &=\sigma\rhe_1 \cos(2 \pi f t)  \,,
\label{eq:strain_rheo_def1bis}\\
\gamma\rhe(t) & = \sum_{k\ge 1} \gamma_k\rhe \cos \left ( 2 \pi k f t + \phi_k\rhe \right ) \,.
\label{eq:gamma_rheo_def1}
\end{align}

Note that one crucial feature of the LORE technique lies in the synchronization of  ultrasonic imaging and rheological data acquisition, which allows us to determine the phase lags $\phi_k\usv (r)$ and thus get accurate measurements of $G'_{\rm loc}(r)$ and $G''_{\rm loc}(r)$. In the present case, synchronization was achieved with a precision of 4~ms by recording the rheological measurements sampled at 250~Hz into an auxiliary file that also stores time stamps for $\Gamma(t)$ and $\Omega(t)$. When ultrasonic acquisition is started, the last time stamp in this file, $t_0$, is retrieved and used as a reference time for both ultrasonic and rheological data [see Fig.~\ref{fig:setup}(b)].


\begin{figure*}
	\centering
	\includegraphics{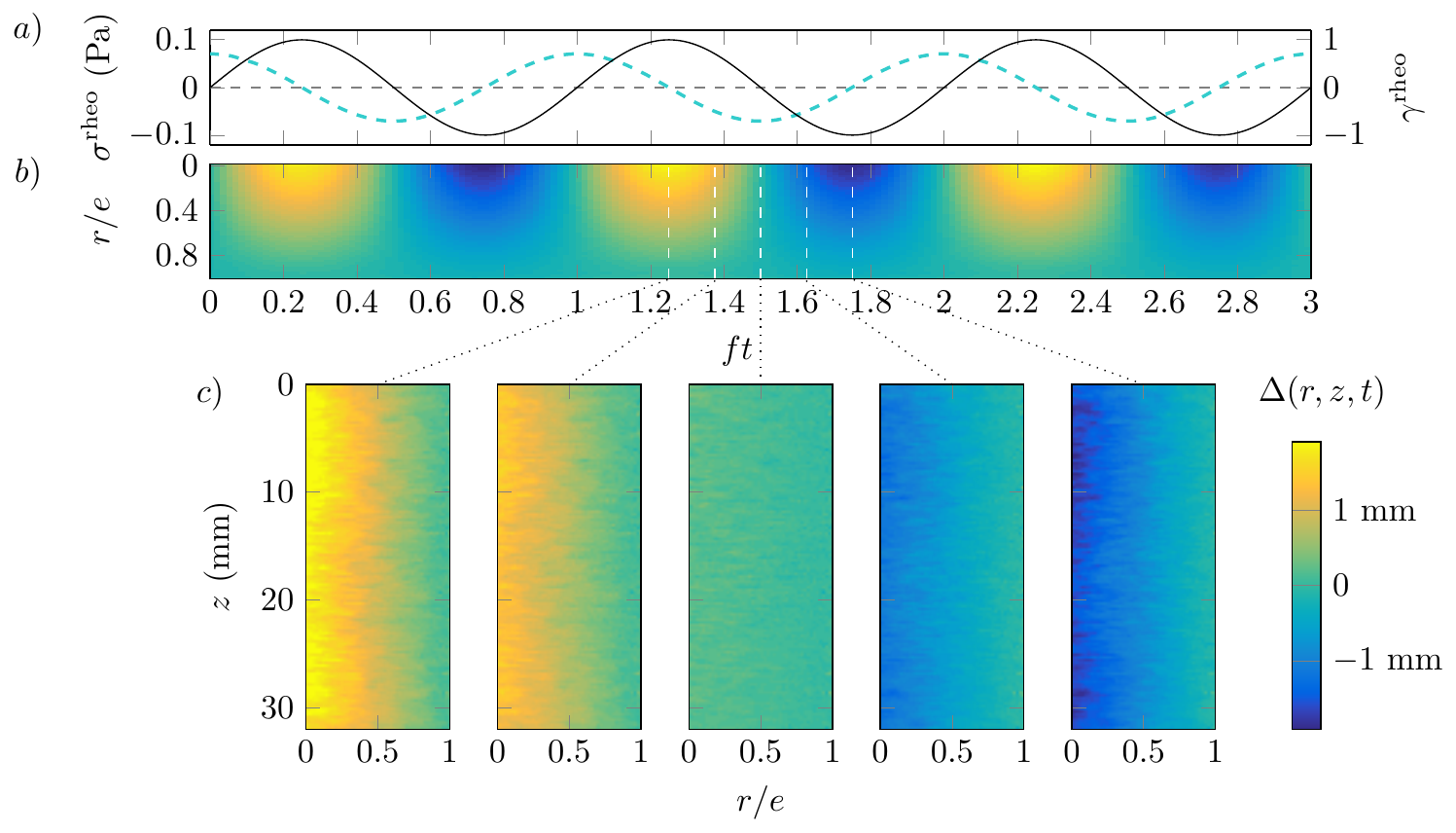}
    \caption{Visualization of the deformation field using ultrasonic imaging during an oscillatory stress experiment in a Newtonian fluid (\textit{Sample 1}). (a)~Stress input $\sigma\rhe(t)$ (blue dashed line) and strain response $\gamma\rhe(t)$ (black line) as recorded by the rheometer. The stress amplitude is $\sigma\rhe=71$~mPa and the corresponding fundamental strain amplitude is $\gamma_1\rhe\simeq 0.99$. (b)~Spatiotemporal diagram of the local displacement averaged along the $z$-direction  $\Delta\usv(r, t)$ as a function of $ft$ and $r/e$. (c)~Snapshots of the local displacement $\Delta\usv(r, z, t)$ in the entire gap taken at times corresponding to $ft=1.25$, 1.375, 1.5, 1.625 and 1.75 from left to right and indicated by dashed lines in (b). Each snapshot results from a moving average on 10 cross-correlations of two successive speckle images separated by 1/60~s. See also Supplemental Movie~1.}
   \label{fig:z_homog}
\end{figure*}

\subsection{Sample preparation}
\label{samples}

In order to benchmark the LORE technique, we focus on four different samples.

\textit{Sample 1} is a mixture of $30\%$ wt. UCON oil (Dow Chemical, 75-H-90,000) and $70 \%$ wt. deionized water which is considered as a purely linear, Newtonian fluid. 

\textit{Sample 2} is a protein gel obtained by the slow acidification of a $6\%$ wt. dispersion of sodium caseinate (Firmenich) in deionized water with $6\%$ wt. glucono-$\delta$-lactone (GDL) (Firmenich)~\cite{Leocmach2014}. This gel is known not to present any wall slip in the present Taylor-Couette geometry~\cite{Leocmach2014, Leocmach2015}. Small oscillations are performed to probe the relative magnitudes of $G'$ and $G''$ during gelation. Combined rheology and ultrasonic imaging are performed after the gelation and once $G'$ and $G''$ 
have reached a steady state. The sample can then be considered as a homogeneous quasi-Hookean soft solid.

\textit{Sample 3} is a two-layer protein gel composed of an outer layer of 1 mm of a concentrated protein gel ($9\%$ wt. sodium caseinate powder acidified with $9\%$ wt. GDL) and an inner layer of 1 mm of a less concentrated protein gel ($5.5\%$ wt. sodium caseinate powder acidified with $5.5\%$ wt. GDL). The outer gel is formed by pouring the yet-to-gel mixture in the rheometer equipped with a rotor of radius $R_{\rm i} = 24$~mm coated with silicone grease. After gelation of the outer layer, the rotor is carefully lifted up at a velocity of $30~\mu$m$\cdot$s$^{-1}$. We then install the usual smaller rotor of radius $R_{\rm i} = 23$~mm and pour the second dispersion that forms the inner gel layer. This process results in a two-layer gel with built-in heterogeneous elastic properties along the radial direction, the part near the rotor being softer than the outer part close to the stator.

\textit{Sample 4} is composed of $2\%$ wt. select Agar (Sigma) in deionized water. After pouring the hot ($\simeq 80^\circ$C) mixture in the Taylor-Couette cell, we wait for temperature equilibration and gelation with the same procedure as for \textit{Sample 2}. \textit{Sample 4}  can also be considered as a homogeneous quasi-Hookean soft solid but, contrary to \textit{Sample 2}, it easily slips at the walls of the Taylor-Couette device.

\section{Results}

\subsection{Homogeneous Newtonian fluid and elastic soft solid}
\label{newtonian}

\begin{figure*}
	\centering
    \includegraphics{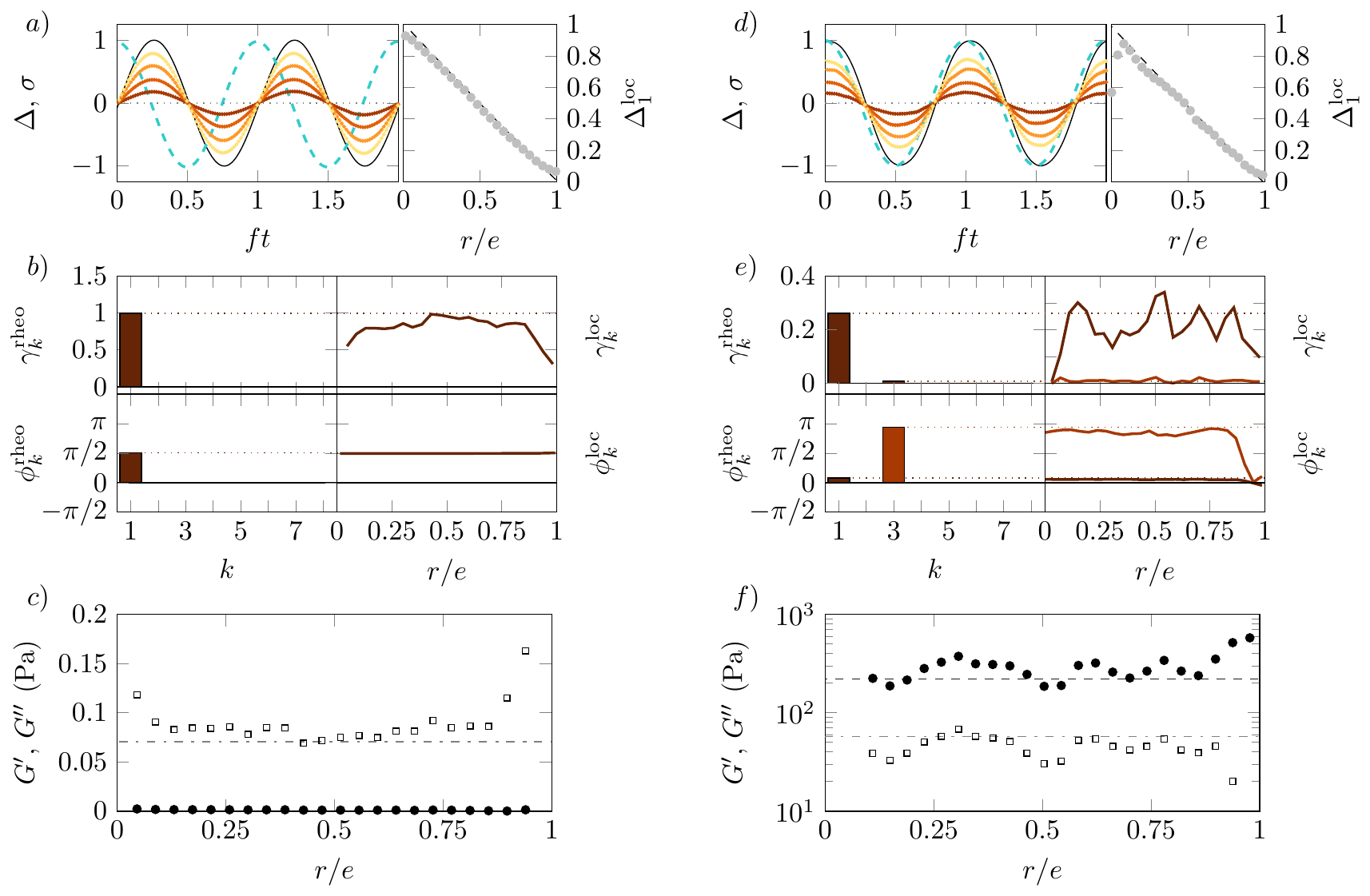}
     \caption{Probing a homogeneous Newtonian fluid (\textit{Sample 1}, left panel) and a homogeneous quasi-Hookean soft solid (\textit{Sample 2}, right panel) with LORE. (a,d)~Left: local ultrasonic displacements $\Delta\usv(r, t)$ (linear color scale from brown at the stator to yellow at the rotor) and rotor displacement $\Delta\rhe(t)=R_{\rm i}\dot{\Omega}$ (black line) in response to an oscillatory stress $\sigma\rhe(t)$ (blue dashed line) as a function of the normalized time $ft$. For \textit{Sample 1}, $\sigma\rhe_1=71$~mPa and $\gamma_1\rhe\simeq 0.99$. For \textit{Sample 2}, $\sigma\rhe_1$= 60 Pa and $\gamma_1\rhe\simeq 0.26$. Right: the amplitude $\Delta_1\usv$ of the fundamental Fourier mode of $\Delta\usv(r, t)$ as a function of $r/e$. The dashed line shows the theoretical profile for a linear homogeneous strain field in Taylor-Couette geometry. (b,e)~Fourier coefficients $\gamma\rhe_k$ and $\phi\rhe_k$ of the rheological strain and phase (left panels) and their local counterparts $\gamma\usv_k$ and $\phi\usv_k$ as a function of $r/e$ (right panels). Colors code for the Fourier mode $k$. (c,f)~Comparison between the viscoelastic moduli obtained from LORE $G'_{\rm loc}(r)$ ($\bullet$) and $G''_{\rm loc}(r)$ (${\small \square}$) and their rheology counterparts $G'_{\rm rheo}$ (dashed line) and $G''_{\rm rheo}$ (dash-dotted line). All measurements displayed in (a) and (d) are normalized by the maximum of the corresponding signal.}
    \label{fig:ucon}
\end{figure*}

\textit{Sample 1}, a Newtonian fluid, and \textit{Sample 2}, a quasi-Hookean soft solid, are both spatially homogeneous materials. Figure~\ref{fig:z_homog} shows that under oscillatory stress, the local deformation in \textit{Sample 1} is in phase-quadrature relative to the sinusoidal stress input $\sigma\rhe(t)$ and behaves homogeneously in the entire gap: as expected, up to experimental noise, the local displacement $\Delta\usv(r, z, t)$ is invariant by translation along the $z$-direction and shows a constant gradient across the gap [see also Fig.~\ref{fig:ucon}(a)]. Displacement maps recorded in \textit{Sample 2} (not shown) are fully similar to Fig.~\ref{fig:z_homog} except that the strain response is in phase with $\sigma\rhe(t)$ [see also Fig.~\ref{fig:ucon}(d)]. Therefore, both materials can be easily and correctly characterized by classical rheological measurements. In the following we use those two samples to benchmark the LORE technique. We also take advantage of the $z$-invariance to average $\Delta\usv$ over the 128 measurement lines along the $z$-direction and thus significantly improve the statistics.

Figure~\ref{fig:ucon}(a) shows the $z$-averaged oscillatory displacement $\Delta\usv(r, t)$ of \textit{Sample 1} measured with ultrasonic imaging in response to an oscillatory stress $\sigma\rhe(t)$ and confirms that these signals are in phase quadrature whatever the position across the gap. Moreover the fundamental mode $\Delta_1\usv(r)$ of the local displacement decreases linearly from the rotor to the stator showing that the sample deformation is homogeneous throughout the gap. The Fourier decomposition of both the global and local strains ascertain more quantitatively the purely viscous nature of \textit{Sample 1} [Fig.~\ref{fig:ucon}(b)]. Indeed, the sample responds harmonically to the stress imposed by the rheometer, all Fourier modes for $k \geq 2$ are negligible, and both the amplitude $\gamma_1\usv(r)$ and the phase  $\phi_1\usv(r)=\pi/2=\phi_1\rhe$ are constant throughout the gap, except for edge effects near the cell walls. 

Finally, the local viscous modulus $G''_{\rm loc}(r)$, computed using Eq.\eqref{eq:G_usv}, is independent of $r$ and matches very well the value provided by the rheometer, $G''_{\rm rheo} = 70$~mPa [Fig.~\ref{fig:ucon}(c)], which corresponds to a fluid of viscosity $\eta = 0.11$~Pa$\cdot$s. The fact that $\gamma_1\usv(r)$ and thus $G''_{\rm loc}(r)$ are space-independent clearly points to a laminar flow consistent with the viscometric assumption used to process global rheological data. However, as recalled e.g. in \cite{Gallot2013,Fardin2014}, one should keep in mind that if the stress amplitude is increased above the onset of inertial or elastic instabilities, secondary flows complicate the picture and may invalidate rheological measurements, therefore making local measurements such as LORE unavoidable. Also note that the presence of the walls leads to spurious static echoes in the ultrasonic speckle signals, which may be difficult to fully dismiss in the data processing \cite{Gallot2013}. This typically leads to underestimating (resp. overestimating) the local displacement close to the rotor (resp. stator) so that, at both walls, the local modulus is generally overestimated. In the data of Fig.~\ref{fig:ucon}, such artifacts extend over roughly $150~\mu$m from the walls. 

\textit{Sample 2}, a casein gel, is known to behave as a homogeneous quasi-Hookean soft solid sticking to the rheometer walls up to strains of about unity~\cite{Leocmach2014}. Figure~\ref{fig:ucon}(d,e) indeed shows that $\Delta\usv(r, t)$ is sinusoidal, proportional to $\sigma\rhe(t)$ and almost in phase with $\sigma\rhe(t)$ in the entire gap, that its amplitude $\Delta_1\usv(r)$ decreases linearly from the rotor to the stator, and that the local strain matches the strain measured by the rheometer. In Fig.~\ref{fig:ucon}(f), we observe that both local viscoelastic moduli, $G'_{\rm loc}(r)$ and $G''_{\rm loc}(r)$, are constant in the entire gap and match the global measurements, $G'_{\rm rheo} = 200$~Pa and $G''_{\rm rheo} = 60$~Pa remarkably well. The non-negligible viscous component $G''$ explains the presence of a slight phase shift between the stress input and the strain response and justifies the term of {\it quasi}-Hookean soft solid. Moreover we note that a small third harmonic ($k=3$) is detected consistently both in rheological and in ultrasonic data, which signals a weakly nonlinear response for $\gamma_1\rhe\simeq 0.26$. Here again, the estimates of $G'_{\rm loc}(r)$ and $G''_{\rm loc}(r)$ suffer from artifacts close to the walls that are inherent to the echography technique.

Overall, we confirm with \textit{Sample 1} and \textit{Sample 2}, respectively a purely viscous fluid and a quasi-Hookean soft solid, that LORE gives access to spatially-resolved measurements of the viscoelastic moduli within a 2-mm gap.

\subsection{LORE in a spatially heterogeneous soft solid}
\label{hetero}

\begin{figure}
	\centering
    \includegraphics{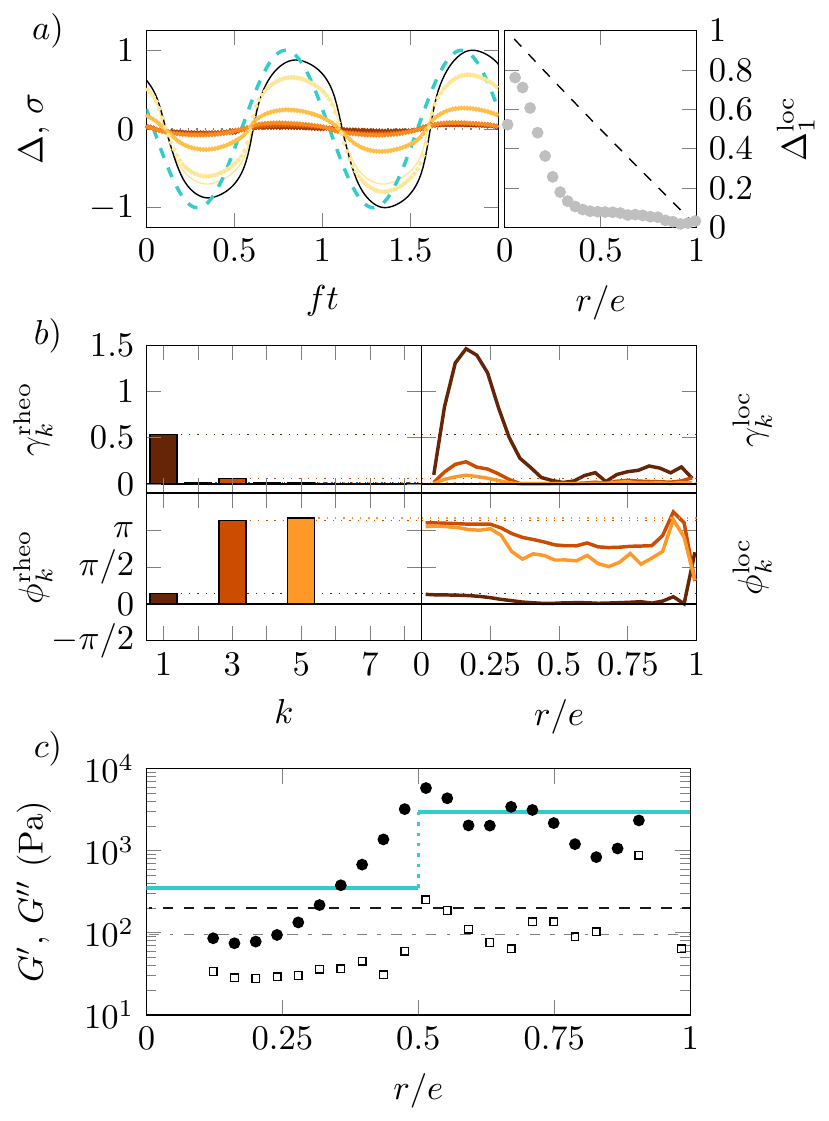}
      \caption{Probing a spatially heterogeneous soft solid (\textit{Sample 3}) with LORE for $\sigma\rhe=119$~Pa and $\gamma\rhe=0.52$. Same legend as in Fig.~\ref{fig:ucon}. For comparison with local measurements in the two-layer casein gel, blue lines in (c) indicate the elastic moduli $G'_{\rm rheo}$ measured independently in two homogeneous casein gels respectively made of 9~\%~wt. sodium caseinate and 9~\%~wt. GDL (left) and made of 5.5~\%~wt. sodium caseinate and 5.5~\%~wt. GDL (right).}
      \label{fig:doubleyaourt}
\end{figure}

\textit{Sample 3} is a heterogeneous protein gel made of a soft inner layer surrounded by a much stronger outer layer. Figure~\ref{fig:doubleyaourt}(a) shows the oscillatory deformation $\Delta\usv(r, t)$ measured with LORE under an oscillatory stress $\sigma\rhe=119$~Pa. In contrast to Fig.~\ref{fig:ucon}(d), the deformation does not decrease linearly with $r/e$. As a consequence, as shown in Fig.~\ref{fig:doubleyaourt}(b), the local strain does not match the strain measured by the rheometer: the gel is much more deformed near the rotor than it is at the stator. This is  consistent with the fact that \textit{Sample 3} is softer near the rotor than near the stator.

In Fig.~\ref{fig:doubleyaourt}(c), we confirm that the evolution of the local viscoelastic moduli is consistent with the local composition of the gel. An outer region can be defined for $r/e>0.5$ where the local values of $G'_{\rm loc}(r)$ fall in the range 2--4~kPa, in good agreement with the global $G'_{\rm rheo}=2.95$~kPa value of a single-layer gel made of 9~\% wt. casein and 9~\% wt. GDL measured independently for similar strain amplitudes [blue lines in Fig.~\ref{fig:doubleyaourt}(c)]. For $r/e<0.5$, the local elastic modulus progressively decreases in the softer inner region down to $G'_{\rm loc}(r)\simeq 80$ Pa at the rotor. For comparison a single-layer gel made of 5.5~\% wt. casein and 5.5~\% wt. GDL has an elastic modulus $G'_{\rm rheo} = 350$~Pa.

In summary, Fig.~\ref{fig:doubleyaourt} shows that the weakest part of the material, which extends over about half the gap, absorbs most of the deformation. The strain close to the rotor is much larger than the global strain of 52\% and reaches roughly 145\% locally, which lies deep in the nonlinear regime~\cite{Leocmach2014}. Accordingly, the presence of odd Fourier Modes up to $k=5$ is reported close to the rotor in the ultrasonic data as well as in the global rheological data (Fig.~\ref{fig:doubleyaourt}b). It important to note that, in the present case of a strongly heterogeneous material, the global viscoelastic measurements $G'_{\rm rheo}$ and $G''_{\rm rheo}$ (dashed and dash-dotted lines in Fig.~\ref{fig:doubleyaourt}c) are off the true local values $G'_{\rm loc}(r)$ and $G''_{\rm loc}(r)$ by up to one order of magnitude.

\subsection{Detection of wall slip through LORE}
\label{slip}

\begin{figure}[h]
	\centering
    \includegraphics{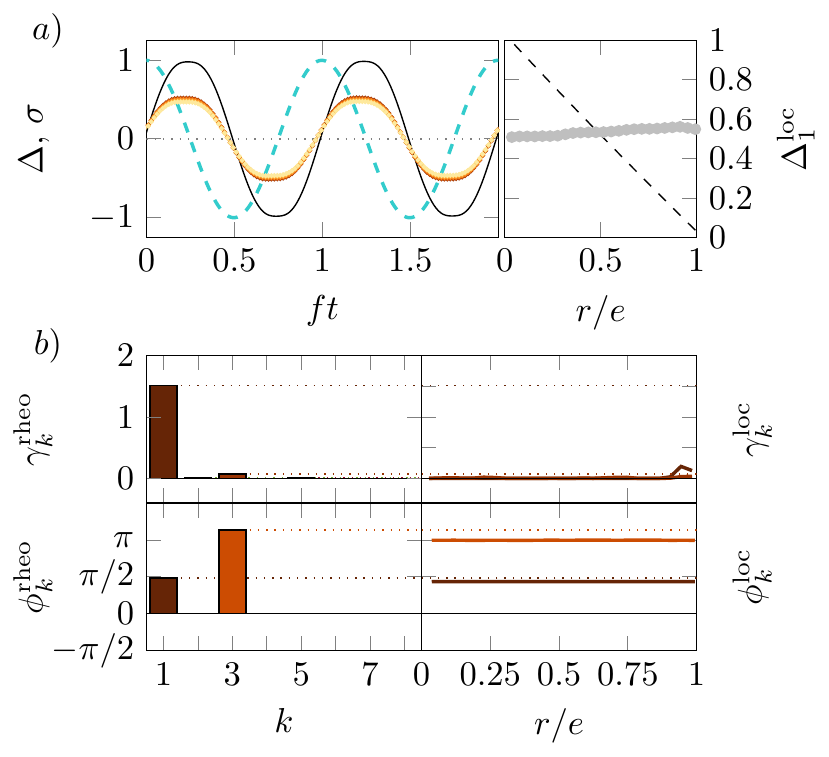}
    \caption{Probing a soft solid slipping at both walls (\textit{Sample 4}) with LORE for $\sigma\rhe = 11.7$~Pa and $\gamma\rhe= 1.44$. Same legend as in Fig.~\ref{fig:ucon}.}
    \label{fig:Agar}
\end{figure}

As recalled in the introduction, apparent wall slip is a very frequent yet still poorly understood phenomenon which seriously complicates the analysis of standard rheological data. In this section we show that LORE proves very useful to detect wall slip under oscillatory shear in the case of an agar gel (\textit{Sample 4}). Such a soft solid is known not to adhere to the smooth PMMA rotor and stator under shear but rather to form thin lubrication layers at both walls due to syneresis, i.e.  the expulsion of water through the gel matrix \cite{Divoux2015}.

From the rheometer point of view, the sample appears as fluidlike: as shown in Fig.~\ref{fig:Agar}(a), the response $\gamma\rhe(t)$ is in phase-quadrature with respect to $\sigma\rhe(t)$. In stark contrast with this apparent fluidlike behaviour, LORE measurements show that the gel remains fully solid in the bulk [see right panel in Fig.~\ref{fig:Agar}(a)]: $\Delta_1\usv(r)$ is constant and non-zero within the entire gap, which means that the gel oscillates as a solid body in Taylor-Couette cell. The amplitude of this solid-body motion is half that of the rotor. Such an oscillatory pluglike flow corresponds to a situation of {\it total wall slip}, in the sense that all the strain applied by the rheometer is actually located in the lubricating layers at both walls while the local strain in the bulk material is effectively zero.

As a consequence, the Fourier analyses of the strain measured by rheology and by ultrasonic imaging are totally different (Fig.~\ref{fig:Agar}b). While rheological measurements point to apparent fluidlike behavior ($\phi\rhe_1=\pi/2$), LORE provides clear evidence for the complete absence of local deformation in the bulk of the sample. As the sample is not sheared in the bulk, it is not possible to measure the local viscoelastic moduli. In this particular case of total slip, we note the emergence of a third-order Fourier mode with no significant second-order mode ($\gamma_2\rhe / \gamma_1\rhe \simeq 2.8 \times 10^{-3}$). In the past literature, the appearance of even harmonics in LAOS experiments has been attributed to slip phenomena \cite{Graham1995,Reimers1996,Klein2007,Guo2011,Hyun2011} while some models have shown that wall slip is not a necessary condition for even harmonics~\cite{Atalik2004}. The LORE measurements of Fig.~\ref{fig:Agar} show that total wall slip alone is also not sufficient to produce even harmonics in the rheological response.

\section{Conclusion}

We have described and tested Local Oscillatory Rheology from Echography (LORE), a new technique based upon the synchronization of high-frequency ultrasonic imaging and oscillatory shear rheometry. We have shown that LORE allows one to access the local viscoelastic moduli and the harmonic content of the local displacement response of soft materials under both linear and nonlinear oscillations. The present paper has been devoted to benchmarking LORE first on homogeneous materials, namely a Newtonian fluid and a quasi-Hookean soft solid and then on a spatially heterogeneous gel as well as on a slipping gel. In the first two cases, LORE provides a direct check that, in homogeneous fluids and solids --and as long as linearity prevails (i.e. in the absence of secondary flows in fluids and of nonlinear effects such as fractures or shear bands in soft solids)--, strain is evenly distributed across the sample and local measurements recover the same values as global rheology. In the two latter cases,  standard rheological estimations are misleading due to the heterogeneity of the sample or to wall slip. There, LORE yields crucial insights into the local dynamics under an oscillatory shear stress by giving access to spatially-resolved $G'$ and $G''$ measurements and/or slip velocities at the cell walls.

As far as further applications are concerned, LORE shall definitely lead to refined insights into the oscillatory response of a wide variety of soft materials with huge industrial importance, ranging from food systems~\cite{Mezzenga2005, Gibaud2012}, such as wheat~\cite{Boire2015}, soybean~\cite{Saio1969} or casein ~\cite{Leocmach2014}, to ``green materials''~\cite{Mohanty2002} like latex~\cite{Oliveira2015} or cellulose~\cite{Martoia2015}. LORE could also help to optimize industrial processes for transiently heterogeneous materials, including hardening concrete~\cite{Roussel2012} or kneaded dough~\cite{Kontogiorgos2014}, as well as materials that are intrinsically submitted to a heterogeneous external field that controls their mechanical properties, such as pipe flows, temperature gradients~\cite{Singh2001,Huang2003} or oxygen concentration promoting heterogeneous polymerization like in dental resin~\cite{Gauthier2005}.

From a more fundamental point of view, yielding~\cite{Bonn2015}, strain hardening~\cite{Gisler1999} and shear thickening~\cite{Brown2010a} are some of the many complex phenomena that could benefit from LORE. For example, during the yielding transition induced by LAOS, local restructuration has been probed using light scattering~\cite{Petekidis2003}, confocal microscopy~\cite{Knowlton2014} or high-frequency ultrasonic echography~\cite{Gibaud2010, Perge2014, Gibaud2015arXiv} but these previous works on yielding dynamics have essentially been limited to stroboscopic measurements from one cycle to the other. With LORE, it becomes possible to map the entire spatiotemporal displacements of the material within a single stress oscillation. Close to the yield point, the response of a soft solid to LAOS, which becomes highly nonlinear and heterogeneous, could be characterized on length scales of a few tens of microns. Extending nonlinear analyses, such as Lissajous-Bowditch representations, Fourier decomposition as used in the present work or more advanced projection techniques~\cite{Hyun2011} to {\it local} measurements is now within reach. Future LORE measurements will therefore help to better understand the intracycle material response and to capture, quantify and predict the rupture of soft solids.

\section*{Acknowledgments}
The authors thank T.~Divoux, B.~Keshavarz and G.~McKinley for fruitful discussions. This work was funded by the Institut Universitaire de France and by the European Research Council under the European Union's Seventh Framework Programme (FP7/2007-2013) / ERC grant agreement No.~258803. 

\bibliography{biblio}

\end{document}